\documentclass[a4paper,11pt]{article}
\usepackage{pos}
\usepackage{bbold}

\newcommand{\la}[1]{\label{#1}}
\newcommand{\ba}{\begin{eqnarray}}
\newcommand{\ea}{\end{eqnarray}}
\newcommand{\bn}{\begin{eqnarray*}}
\newcommand{\en}{\end{eqnarray*}}
\newcommand{\fig}{Fig.~}
\newcommand{\eq}{Eq.~}
\newcommand{\se}{Sec.~}

\renewcommand{\nr}[1]{(\ref{#1})}
\newcommand{\ep}{\varepsilon}
\newcommand{\ccdot}{\!\cdot\!}
\newcommand{\rmi}[1]{{\mbox{\scriptsize #1}}}
\newcommand{\Nc}{N_\rmi{c}}
\newcommand{\Nf}{N_\rmi{f}}
\newcommand{\gE}{g_\rmi{E}}
\newcommand{\mE}{m_\rmi{E}}
\newcommand{\lambdaE}{\lambda_\rmi{E}}
\newcommand{\mel}{m_\rmi{el}}
\renewcommand{\vec}[1]{{\bf #1}}
\newcommand{\Tint}[1]{{\hbox{$\sum$}\!\!\!\!\!\!\int\,}_{\!\!\!\!\!\!\raise-0.2ex\hbox{$\scriptstyle{#1}$}}}
\newcommand{\ci}{{\cal I}}
\newcommand{\tK}{\tilde K}
\newcommand{\code}[1]{{\fontfamily{qcr}\selectfont #1}}
\newcommand{\Ca}{C_\rmi{A}}
\newcommand{\Cf}{C_\rmi{F}}

\title{Debye screening mass in hot QCD at three loops: Canonical form of the integrand}

\author*[a]{York Schr\"{o}der}
\author[a,b]{Miguel Vaez}

\affiliation[a]{Centro de Ciencias Exactas, Departamento de Ciencias B\'asicas, Universidad del B\'io-B\'io, 
\\Avenida Andr\'es Bello 720, Chill\'an, Chile}

\affiliation[b]{Departamento de F\'isica, Facultad de Ciencias, Universidad del B\'io-B\'io, 
\\Avenida Collao 1202, Concepción, Chile}

\emailAdd{yschroder@ubiobio.cl}
\emailAdd{miguel.vaez2401@alumnos.ubiobio.cl}

\abstract{The Debye screening mass is a fundamental parameter characterizing the screening of chromo-electric fields in a hot quark-gluon plasma. It plays an important role in thermal field theory, entering the description of screening lengths, transport phenomena, heavy-particle interactions in the early Universe, and the dimensional reduction program that connects thermal QCD to effective three-dimensional theories. In this contribution, we revisit the reduction strategy that serves to organize the large number of terms that contribute to the screening mass at the three-loop level, applying recent advances based on canonical forms of thermal (sum-) integrands. We emphasize the role of integration variable shifts, and discuss how canonical representations of thermal integrands may simplify future high-order calculations in hot QCD.}

\FullConference{Loops and Legs in Quantum Field Theory (LL2026)\\
12-17, April, 2026\\
Bayreuth, Germany\\}

\begin{document}
\maketitle


\section{Introduction}

Thermal QCD at high temperatures $T$ is characterized by a hierarchy of dynamically generated momentum scales
\ba
2\pi T \gg gT \gg g^2T \;,
\ea
where $g=\sqrt{4\pi\alpha_s}$ is the strong gauge coupling constant that is decreasing at high $T$ due to asymptotic freedom.
The interplay of these scales is responsible for both the richness and the difficulty of finite-temperature gauge theories (for a modern textbook-style treatment and relevant references, see e.g.\ \cite{Laine:2016hma}). While the so-called hard momentum scale $2\pi T$ can be treated using conventional perturbation theory, the other two (soft and ultrasoft) scales generate infrared complications and eventually lead to the breakdown of ordinary perturbative expansions \cite{Appelquist:1981vg,Linde:1980ts}, necessitating a reorganized weak-coupling expansion strategy. A systematic option is the framework of dimensionally reduced effective theories, to which we will adhere here.

One key quantity associated with the soft sector is the Debye screening mass. In analogy with electromagnetic plasmas, where the Coulomb force between two test charges weakens due to their interactions with the plasma particles leading to a Yukawa-type potential $\propto -\alpha \exp(-\mE r)/r$, it parametrizes the screening of static chromo-electric gluon fields in the quark-gluon plasma. Beyond leading order, however, its definition becomes somewhat subtle. Gauge-dependent definitions based on the static gluon self-energy are insufficient, while gauge-invariant Debye screening masses become intrinsically non-perturbative quantities \cite{Rebhan:1993az,Arnold:1995bh}. A particularly useful alternative emerges within the framework of effective field theory, where the Debye mass $\mE$ appears as a matching coefficient of the dimensionally reduced effective theory known as Electrostatic QCD (EQCD) \cite{Braaten:1995cm,Kajantie:1997tt}.

At leading order one finds for a SU($\Nc$) gauge theory with $\Nf$ massless fermions \cite{Kapusta:1979fh}
\ba
\mE^2 &=&
g^2 T^2
\Big( \frac{\Nc}{3}+\frac{\Nf}{6} \Big) \;,
\ea
but higher-order corrections rapidly become non-trivial to evaluate. Early work established the dimensional-reduction framework and the effective-theory description of hot QCD \cite{Braaten:1995cm,Kajantie:1997tt}. Subsequent investigations incorporated quark-mass threshold effects \cite{Laine:2006cp}, developed the reduction machinery required for three-loop matching coefficients \cite{Moller:2012chx}, completed the three-loop Debye-mass calculation in pure Yang-Mills theory \cite{Ghisoiu:2015uza}, and finally extended the Debye mass to two-loop order in full QCD including finite quark masses \cite{Laine:2019uua}.

The purpose of this contribution is to revisit these developments and to discuss recent progress based on canonical forms of thermal integrands \cite{Navarrete:2024ruu}. These methods provide a novel organizational principle for multiloop finite-temperature calculations and may significantly simplify future computations beyond three loops.


\section{EQCD and matching coefficients}

Within the imaginary time formulation of thermal field theory, the infrared structure of hot QCD can be organized systematically through dimensional reduction from $D=4-2\ep$ to $d=3-2\ep$ dimensions \cite{Braaten:1995cm,Kajantie:1997tt}. Integrating out the hard Matsubara modes leads to a purely bosonic three-dimensional effective theory, EQCD, whose Euclidean Lagrangian reads
\ba \la{eq:eqcd}
{\cal L}_\rmi{EQCD} &=&
-\frac1{2\gE^2}\,{\rm Tr}\,[D_i,D_j]^2
+{\rm Tr}[D_i,A_0]^2
+\mE^2\,{\rm Tr}\,A_0^2
+\lambdaE^{(1)} \big({\rm Tr}A_0^2 \big)^2
+\lambdaE^{(2)}\,{\rm Tr}\,A_0^4
+\dots \;,\quad
\ea
with covariant derivative $D_i=\mathbb{1}\partial_i-i\gE A_i$ and $A=A^aT^a$, and where the dots stand for the infinite tower of higher-order operators that are induced when integrating out the hard modes but do not contribute here.
Its parameters $\{\gE^2,\,\mE^2,\,\lambdaE^{(1)},\,\lambdaE^{(2)},\,\dots\}$ encode the effects of the hard thermal scale and are determined as functions of the parameters $\{g^2,\,T,\,\Nc,\,\Nf,\,m_\rmi{f},\,\mu_\rmi{f}\}$ of full QCD through matching calculations that ensure that both theories describe the same long-range physics.

A convenient condition to determine e.g.\ the EQCD mass parameter $\mE$ is to require propagator poles to match. In full QCD, the electric screening mass is obtained from the pole of the static ($P_0=0$) temporal momentum-space gluon propagator with the on-shell condition $\vec p^2=-\mel^2$,
\ba
0 &=& P_0^2+\vec p^2+\Pi_{00}(P_0,\vec p) \Big|_{P_0=0,\;\vec p^2=-\mel^2} \;,
\ea
where we have written the gluon self-energy of full QCD as $\Pi_{\mu\nu}^{ab}(P)=\delta^{ab}\Pi_{\mu\nu}(P_0,\vec p)$. Performing a loop-expansion as well as a Taylor expansion around soft external $|\vec p|\propto gT$ of the self-energy as
\ba \la{eq:taylor}
\Pi_{00}(0,\vec p) &=&
\sum_{n=1}^{\infty} g^{2n}\,\Pi_\rmi{En}(\vec p^2) 
\;=\; 
\sum_{n=1}^{\infty} g^{2n}\,\Big\{
\Pi_\rmi{En}(0) 
+\vec p^2\,\Pi'_\rmi{En}(0) 
+\big(\vec p^2\big)^2\,\Pi''_\rmi{En}(0) 
+\dots\Big\}\;,
\ea
one obtains the screening mass as a strict perturbative series in terms of moments of the temporal gluon self-energy (dropping the zero arguments for brevity) 
\ba \la{eq:mE}
\mel^2 &=&
g^2\,\Pi_\rmi{E1}
+g^4 \Big[ \Pi_\rmi{E2} - \Pi'_\rmi{E1}\Pi_\rmi{E1} \Big]
\nonumber\\
&&{}+g^6\Big[
\Pi_\rmi{E3}
-\Pi'_\rmi{E1}\Pi_\rmi{E2}
-\Pi'_\rmi{E2}\Pi_\rmi{E1}
+\Pi''_\rmi{E1}\Pi_\rmi{E1}\Pi_\rmi{E1}
+\Pi'_\rmi{E1}\Pi'_\rmi{E1}\Pi_\rmi{E1}
\Big] +{\cal O}(g^8)\;.
\ea

On the EQCD side, one considers the pole of the 3d adjoint scalar $A_0$ propagator as defined by 
\ba
0&=& \vec p^2+\mE^2+\Pi_\rmi{EQCD}(\vec p)\;\big|_{\vec p^2=-\mel^2} \;.
\ea
Now, within a strict perturbative treatment of EQCD (treating the external momentum $\vec p^2$ as well as the tree-level mass $\mE^2$ as perturbatively small), all resulting 3d vacuum integrals that contribute to the loop expansion of $\Pi_\rmi{EQCD}$ become scale-free and therefore vanish in dimensional regularization. Consequently,
\ba
\mE^2 = \mel^2 \;,
\ea
and the determination of the Debye mass reduces to a computation of moments of thermal self-energies in full QCD according to the right-hand side of \eq\nr{eq:mE}.

The three-loop evaluation of Eq.~(\ref{eq:mE}) generates several hundred Feynman diagrams (cf.\ \fig\ref{fig:diags}) and millions of intermediate sum-integrals. Using integration-by-parts (IBP) identities adapted to finite temperature \cite{Nishimura:2012ee}, these expressions have been reduced to a comparatively small set of master sum-integrals \cite{Moller:2012chx}. A major result of Ref.~\cite{Moller:2012chx} was the demonstration that $\mE^2$ remains explicitly gauge independent after IBP reduction, providing a stringent consistency check of the entire framework.

\begin{figure}
\centering 
\includegraphics[width=1.0\textwidth]{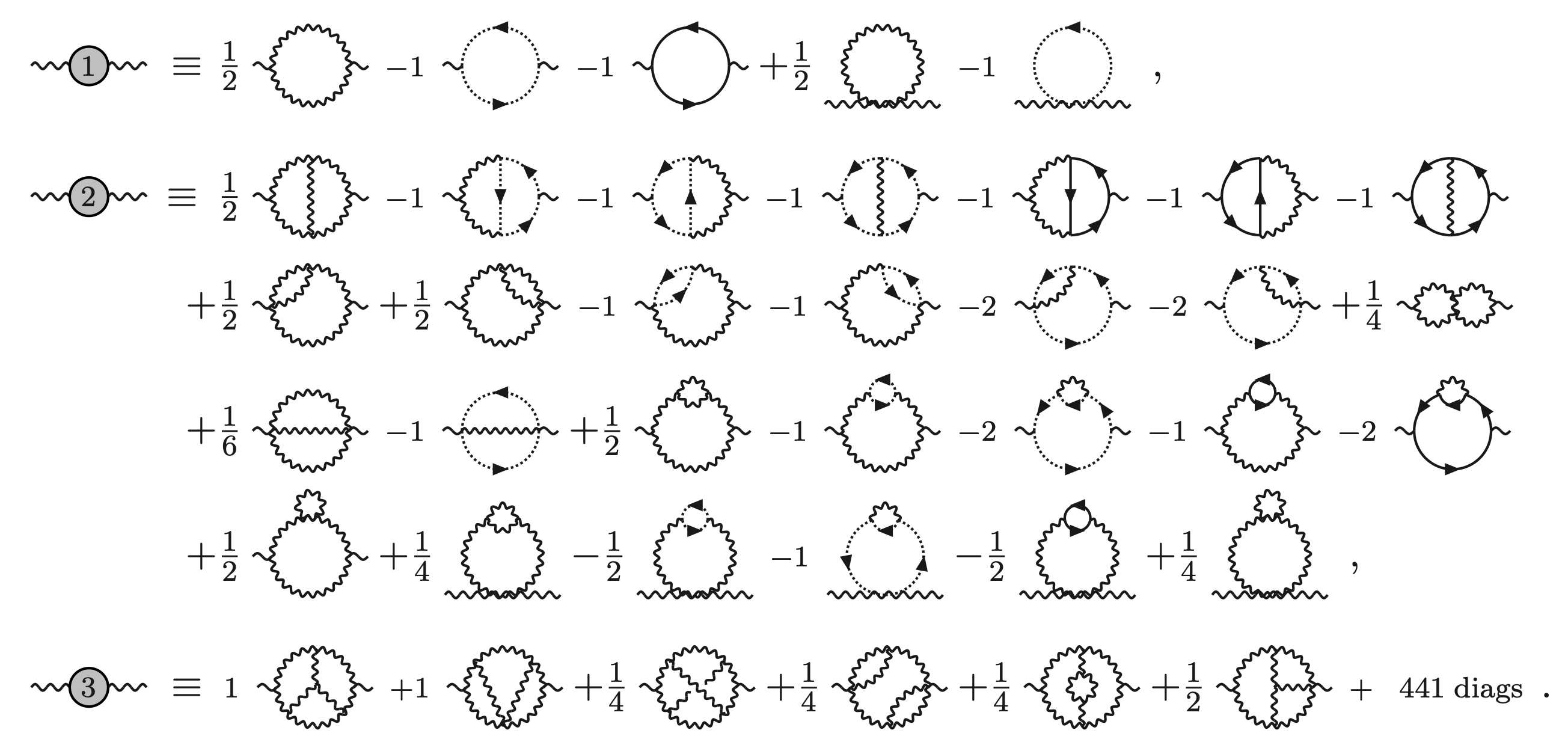}
\caption{\la{fig:diags}The 1-loop, 2-loop and some 3-loop self-energy diagrams in background field gauge.
Wavy lines represent gluons, dotted lines ghosts, and solid lines quarks. Figure taken from \cite{Moller:2012chx}.}
\end{figure}

In the following, we ask the question whether this reduction can already be achieved by simpler means \cite{MVthesis}, namely by applying the integrand canonicalization algorithm that had been successfully applied in the case of the hot QCD pressure \cite{Navarrete:2022adz,Navarrete:2024ruu}, working over rational numbers only and therefore avoiding the $D$-dependent polynomial algebra necessary for the full IBP reduction algorithm.


\section{Sum-integral classification}

In order to evaluate $\mE^2$, it is useful to cast it into a form that is easy to manipulate and simplify algorithmically.
We work in the background field gauge and set all quark masses and chemical potentials to zero here. 
In a first step, we generate the complete set of (connected, one-particle irreducible) self-energy diagrams as shown in \fig\ref{fig:diags} and expand in the external momentum $P$ to the respective orders needed for the moments entering \eq\nr{eq:mE}. 
The expansion can be performed by simply iterating the identity $\frac1{(K+P)^2}=\frac1{K^2}-\frac{2K\ccdot P+P^2}{K^2\,(K+P)^2}$ where $K$ stands for loop momenta, up to the required order of numerator factors $P$ and then nullifying it in the denominators. 

At this stage all (sum-) integrals are of vacuum type, which can be scalarized by applying fully symmetric Lorentz projectors containing the metric $\delta^{\mu\nu}$ and the four-vector $U^\mu=(1,\vec 0)$ that is present in thermal field theory. 
Noting that integrals containing an odd power of $P$ in the numerator vanish due to the integral's symmetry under sign flips of all loop integration momenta, one recovers precisely the expansion indicated in \eq\nr{eq:taylor}, with all coefficients given by scalar vacuum-type sum-integrals that in the 3-loop case can be classified by a lists of integers $(c_1\dots c_3,a_1\dots a_3,s_1\dots s_6)$ as
\ba \la{eq:actors}
\ci(\vec c,\vec a, \vec s) &=& \Tint{K_1} \Tint{K_2} \Tint{K_3} \frac{(U\ccdot \tK_1)^{a_1}\cdots(U\ccdot \tK_3)^{a_3}}{(P_1\ccdot P_1)^{s_1}\cdots (P_{6}\ccdot P_{6})^{s_{6}}} \;.
\ea
Here, positive (negative) indices $s_i\in\mathbb{Z}$ correspond to propagators (numerators), the $a_i\in\mathbb{N}$ are non-negative, and the $c_i\in\{0,1\}$ differentiate bosonic and fermionic Matsubara modes as $\tK_i^\mu=K_i^\mu+c_i \pi T U^\mu$ for $i\in\{1,2,3\}$ where the loop four-momenta are $K_i^\mu=(2n_i\pi T,\vec k_i)$ and we define the sum-integral symbol as
\ba
\Tint{K_i} = T\sum_{n_i\in\mathbb{Z}} \int\frac{{\rm d}^d\vec k_i}{(2\pi)^d} \;.
\ea
In \eq\nr{eq:actors}, we have mapped all vacuum integrals to our preferred 3-loop momentum family constructed from six linear combinations of the three loop momenta $\tK_1^\mu\dots\tK_3^\mu$ given by (omitting Lorentz indices)
\ba \la{eq:family}
\{P_1,\dots,P_6\} &=& \{\tK_1,\tK_2,\tK_3,\tK_1-\tK_2,\tK_1-\tK_3,\tK_2-\tK_3\} \;.
\ea
For completeness, let us mention that we define the lower-loop sum-integrals occurring in \eq\nr{eq:mE} in full analogy, by mapping all momenta onto the one-loop family $\{P_1\}=\{\tK_1\}$ and two-loop family $\{P_1,\dots,P_3\}=\{\tK_1,\tK_2,\tK_1-\tK_2\}$, respectively. 


\section{Sum-integral reduction}

Having expressed all sum-integrals contributing to \eq\nr{eq:mE} in the form of \eq\nr{eq:actors}, we now strive to reduce the number of independent evaluations to a minimum. Indeed, not all of the (tens of thousands of) sum-integrals generated after multiplying out the Feynman rules (for a list in background field gauge, see e.g.\ \cite{MVthesis}; our gluon propagator is $[\delta^{\mu\nu}+\xi K^\mu K^\nu/K^2]/K^2$ with gauge parameter $\xi$) and performing Lorentz-, Dirac- and color-algebra are independent, but related via linear relations of type $0=\sum c\cdot\ci$, where the coefficients $c=c(D)$ can be rational functions in the space-time dimension $D=\delta^{\mu\mu}$. These linear relations are used systematically, in order to eliminate redundancies and reduce the number of sum-integral evaluations, which become non-trivial starting at three loops \cite{Davydychev:2023jto,Gil:2026cqz,Ghisoiu:2012yk}.

As is well known, the key underlying strategy to systematically reduce a large set of linear relations is an ordering relation among the unknowns $\ci$. To this end, we assign a list of integers such as loop number, number of propagators, extra propagator powers, etc.\ to each sum-integral which then allows to compare any two $\ci$ by comparing those lists from the left until an element differs, at which point an ordering like $\ci_1 \prec \ci_2$ is uniquely established.


\subsection{Sector shifts}

One important set of linear relations is based on trivial changes of integration variables, such as e.g.\ $K_1\leftrightarrow K_2$, or indeed any linear transformation $K_i\rightarrow S_{ij}\,K_j$ that can be represented by a square shift matrix $S$ with integer coefficients and $\det S=\pm1$, such that the integral measure remains invariant. Such shifts act on the numerator and denominator structure of the $\ci$ and therefore induce linear relations among the integrands, which can then be exploited given the ordering relation, allowing to systematically eliminate 'harder' for 'simpler' ones. 

We call one such useful class of shifts $S$ 'sector shifts', the name of which will become clear in a moment. 
By 'sector' we refer to a subset of the momentum family \eq\nr{eq:family}, which we label by an integer \code{sectorID} according to the binary representation of the propagator momenta present in the integrand, such as e.g.\ $\{K_1,K_2,K_1-K_3,K_2-K_3\}=\{P_1,P_2,P_5,P_6\}\rightarrow110011_2=51$. Different sectors can correspond to the same underlying graph, such as $51\,\hat{=}\,45\,\hat{=}\,30$. As is well established in the collider physics community, let us note that since isomorphic graphs have the same Symanzik polynomials and such polynomials can be normal ordered by efficient algorithms \cite{Pak:2011xt,Hoff:2015kub}, it is easy to identify shifts $S$ that induce such graph isomorphisms and use the linear relations $0=\ci-(S\circ\ci)$ (where the shift is understood to act under the sum-integral sign) to e.g.\ maximize the \code{sectorID}, i.e.\ choosing 5 unique 3-loop sector representatives (out of $2^6=64$) as depicted in \fig\ref{fig:sectors}.

\begin{figure}
\centering 
\includegraphics[width=1.0\textwidth]{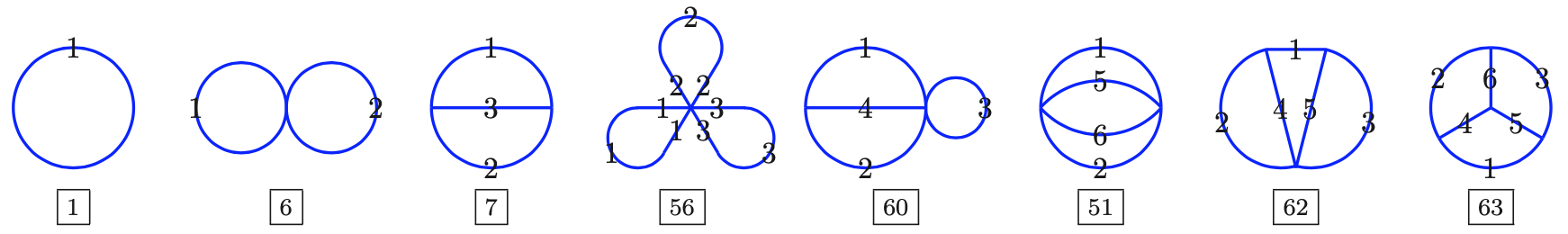}
\caption{\la{fig:sectors}Sector representatives relevant for our calculation. The enumeration of lines follows the momentum families explained around \eq\nr{eq:family}, and the boxed numbers are the corresponding identifiers \code{sectorID}.}
\end{figure}


\subsection{Symmetry shifts}

Another useful class of shifts $S$ are 'symmetry shifts', which do not change the \code{sectorID}.
These induce sector automorphisms, i.e.\ they leave the denominator structure of the integrand of $\ci$ invariant.
As such symmetry shifts can nevertheless change the index order as well as the numerator structure of $\ci$, they serve to further eliminate redundancies among the set of unknown sum-integrals, and we find it most advantageous to use the complete set of such symmetries for each sector. Indeed, since the shift matrices are unimodular they form a group under matrix multiplication, and we need to only record the respective generators. This is highly efficient, as for example just two $S$ generate all 24 symmetry shifts of sector 51. 
Again, we can feed linear relations $0=\ci-(S\circ\ci)$ generated by the symmetry shifts $S$ into our system and systematically eliminate 'hard' sum-integrals guided by the ordering relation. 

Concretely, as we only need to consider shifts up to an overall sign-flip of all integration momenta $K_i$, we find it sufficient to record only three 2-loop generators
\ba
S_1=\left(
\begin{array}{cc}
 1 & 0 \\
 0 & -1 \\
\end{array}
\right)\;,\quad S_2=\left(
\begin{array}{cc}
 0 & 1 \\
 1 & 0 \\
\end{array}
\right)\;,\quad S_3=\left(
\begin{array}{cc}
 1 & 0 \\
 1 & -1 \\
\end{array}
\right)\;.
\ea
The $\{4,6\}$ symmetries of sectors $\{6,7\}$ (cf.\ \fig\ref{fig:sectors}) are then generated by (products of) the shift matrices $\{\{S_1,S_2\},\,\{S_2,S_3\}\}$, respectively. At three loops, we need merely six generators 
\ba
&&S_1=\left(
\begin{array}{ccc}
 0 & 0 & 1 \\
 0 & 1 & 0 \\
 -1 & 0 & 0 \\
\end{array}
\right)\;,\quad S_2=\left(
\begin{array}{ccc}
 0 & 1 & 0 \\
 1 & 0 & 0 \\
 0 & 0 & 1 \\
\end{array}
\right)\;,\quad S_3=\left(
\begin{array}{ccc}
 1 & 0 & 0 \\
 1 & -1 & 0 \\
 0 & 0 & 1 \\
\end{array}
\right)\;,\quad \\&&S_4=\left(
\begin{array}{ccc}
 1 & 0 & 0 \\
 0 & 0 & 1 \\
 0 & 1 & 0 \\
\end{array}
\right)\;,\quad S_5=\left(
\begin{array}{ccc}
 0 & 1 & 0 \\
 0 & 1 & -1 \\
 -1 & 1 & 0 \\
\end{array}
\right)\;,\quad S_6=\left(
\begin{array}{ccc}
 1 & 0 & 0 \\
 0 & -1 & 1 \\
 0 & 0 & 1 \\
\end{array}
\right)\;.
\ea
The $\{24,12,8,24,24\}$ symmetries of sectors $\{56, 60, 62, 63, 51\}$ are then generated by (products of) the shift matrices 
$\{\{S_1,S_2\},\,\{S_2,S_3\},\,\{S_3,S_4\},\,\{S_4,S_5\},\,\{S_5,S_6\}\}$, respectively.

Iteratively applying the sector- as well as symmetry-shifts to our class of sum-integrals, we map all integrands to a canonical form that is invariant under all such changes of integration variables, eliminating an enormous number of redundancies among the $\ci$. This powerful analytic mapping method in fact only needs algebra over the integers, and not over rational functions (in the variable $D$ in our case, such as one would have encountered when employing a full IBP reduction strategy from the outset). 

\begin{table}
\centering
\begin{tabular}{|l||c|c|c|c|c|c|}
\hline
Color/flavor structure & \multicolumn{3}{c|}{$\Ca^2$} & \multicolumn{2}{c|}{$\Nf\,\Ca$} & $\Nf\,\Cf$ \\
\hline
Gauge parameter & $\xi^3$ & $\xi^2$ & $\xi^1$ & $\xi^2$ & $\xi^1$ & $\xi^1$\\
\hline
Number of sum-ints & 53 & 71 & 66 & 20 & 39 & 15 \\
after sector shifts & 25 & 27 & 32 &  8 & 26 & 13 \\
after symmetry shifts &  0 &  0 & 10 &  0 & 10 & 0 \\
after IBP 'light' &  0 &  0 &  0 &  0 &  0 & 0 \\
\hline
\end{tabular}
\caption{\la{table2}Reduction pattern for thermal sum-integrands entering the 2-loop term of \eq\nr{eq:mE}, at different stages of the canonicalization procedure as explained in the main text (see \se\ref{se:ibp} for details on IBP 'light'). Gauge parameter dependence cancels algebraically for all terms, as evidenced by the zero entries in the table.}
\end{table}

In Table~\ref{table2}, we illustrate the reduction in the total number of different sum-integrals that comprise the 2-loop ($g^4$) term of \eq\nr{eq:mE}, for different representative structures as indexed by their color group prefactors and gauge parameter power. We observe cancellation of all gauge parameter dependence after the canonicalization steps. While the complete 2-loop reduction to master sum-integrals could in fact have been performed analytically from the outset by known symbolic recurrence relations \cite{Davydychev:2023jto,Gil:2026cqz}, we take the results as a proof-of-principle of the integrand canonicalization setup, noting that, in contrast to the case of the QCD pressure \cite{Navarrete:2024ruu}, here also some simple $D$-independent IBP relations acting on the factorized sector 6 need to be employed, see \se\ref{se:ibp} for details.

\begin{table}[t]
\centering
\begin{tabular}{|l||c|c|c|c|c|c|c|c|c|c|c|c|}
\hline
color & \multicolumn{2}{c|}{$\Ca^3$} & \multicolumn{2}{c|}{$\Nf\,\Ca^2$} & \multicolumn{2}{c|}{$\Nf\,\Ca\,\Cf$} & \multicolumn{2}{c|}{$\Nf\,\Cf^2$} & \multicolumn{2}{c|}{$\Nf^2\,\Ca$} & \multicolumn{2}{c|}{$\Nf^2\,\Cf$} \\
\hline
gauge & $\xi^5$ & $\xi^2$ & $\xi^4$ & $\xi^2$ & $\xi^3$ & $\xi^1$ & $\xi^2$ & $\xi^1$ & $\xi^2$ & $\xi^1$ & $\xi^2$ & $\xi^1$\\
\hline
\# ints & 3661 & 9314 & 595 & 5471 & 371 & 2409 &292&495 &320&569 &70&220 \\
sector & 805 & 1953 & 114 &  1522 & 105 & 991 &167&235 &111&324 &17&141 \\
symm &  0 &  99 & 0 &  123 & 0 & 20 &0&0 &28&95 &0&0 \\
IBP &  0 &  0 &  0 &  0 &  0 & 0 &0&0 &0&0 &0&0 \\
\hline
\end{tabular}
\caption{\la{table3}Same as Table~\ref{table2}, for some representative structures that occur in the 3-loop term of \eq\nr{eq:mE}. Again, only a 'light' version of IBP is required to achieve full gauge parameter cancellation.}
\end{table}

In Table~\ref{table3}, the same scheme is illustrated for the 3-loop ($g^6$) term of \eq\nr{eq:mE}. Again, we observe a large reduction of terms when applying both types of shift relations to organize the integrand into its canonical form, and gauge parameter independence after adding $D$-independent IBP relations for the factorized sectors 60 and 56 to the linear system.


\subsection{Example}

To show the general idea of applying these shifts in a very simple example, consider the 2-loop shift $S_1={\rm diag}(1,-1)$, which (denoting the massless inverse propagators as $P_n\ccdot P_n=D_n$) acts on the propagator family as $\{D_1,D_2,D_3\}\rightarrow\{D_1,D_2,2D_1+2D_2-D_3\}$. It is therefore a symmetry of sector 6, but not of sector 7.
Applying it to e.g.\ the integral $I(3,2,-1)$ which is an instance of sector 6 with one numerator factor (we abbreviate $\ci(\vec 0,\vec 0,\vec s)\equiv I(\vec s)$ here) produces the identity
\ba
0 &=& \Big\{ S_1-\mathbb{1}\Big\}\circ I(3,2,-1) 
\label{eq:shiftExample}
\;=\; 2I(2,2,0) +2I(3,1,0) -I(3,2,-1) -I(3,2,-1) \;,
\ea 
where both (shift and identity) matrices are understood to act as momentum transformations of the sum-integral's measure.
Eq.~(\ref{eq:shiftExample}) can be solved for its 'most difficult' element according to the ordering relation, which for our conventions produces
\ba
I(3,2,-1) &=& I(2,2,0) + I(3,1,0) \;.
\ea
As a result, we got rid of the numerator factor in this example (in fact obtaining two sum-integrals that simply are products of scalar 1-loop factors), using only algebra over the integers 
and avoiding tensor decomposition.


\subsection{IBP relations}
\la{se:ibp}

Finally, integration-by-parts (IBP) relations provide further linear relations among the $\ci$, in the case at hand with coefficients that are single-variable rational functions in the space-time dimension $D$. The generic strategy and its finite-temperature modification have been documented e.g.\ in \cite{Nishimura:2012ee}, but let us recall that in the thermal setting the derivatives are taken with respect to spatial loop momenta only such that IBP relations are generated via ($i,j\in\{1,2,3\}$)
\ba
0 &=& \partial_{\vec k_i} \ccdot \vec k_j \circ \ci(\vec c,\vec a,\vec s) 
\;=\; \big[(D-1)\delta_{ij}+\vec k_j \ccdot \partial_{\vec k_i}\big] \circ \ci(\vec c,\vec a,\vec s) 
\;,
\ea
where the derivative is understood to act under the integral sign. For each seed integral $\ci$, linear relations are obtained by working out the derivatives and re-writing the result in terms of our generic sum-integral \eq\nr{eq:actors}. 

To postpone working over rational functions as long as possible, we can add a number of $D$-independent IBP relations (e.g.\ the six non-diagonal relations resulting from picking $i\neq j$, and two differences of diagonal relations) to our integer-coefficient linear system. In fact, it turns out that doing this when the seed integral belongs to a factorized sector is sufficient to reveal the gauge-invariant representation in terms of canonical sum-integrals for $\mE^2$, as indicated in the tables. This is what is indicated as 
IBP 'light' in the tables, complementing the symmetrization procedure.

Of course, in order to achieve a truly minimal representation in terms of master sum-integrals also for the gauge-parameter free terms, the final $D$-dependent relations can be added to the system, triggering the rational function algebra that is often presenting a bottleneck in large-scale IBP reductions, causing intermediate relations to grow very large, and potentially generating coefficients with spurious poles $\propto1/(D-4)$ which would require the evaluation of some master sum-integrals to higher orders in $\ep$. There are of course a number of improvements that can be (and have been) applied to remedy all of these points and that have been reported in this conference series, but this is not our main point here.


\section{Conclusions}

The Debye screening mass occupies a central role in the effective-theory description of hot QCD. Through dimensional reduction it emerges as a matching coefficient of EQCD and provides a perturbatively calculable bridge between the hard and soft thermal scales. Over the last two decades substantial progress has been achieved through the development of matching techniques, the inclusion of finite quark-mass effects, the determination of the three-loop Yang--Mills Debye mass, and the extension to realistic QCD with massive quarks.

Recent developments based on canonical forms of thermal integrands suggest an efficient approach to multiloop thermal field theory. By systematically eliminating redundancies before full IBP integral reduction, these methods promise to simplify future calculations and may help extend the perturbative frontier of hot QCD beyond its current limits.

Looking ahead, these methods may play an important role in completing several outstanding perturbative projects in hot QCD, including higher-order matching coefficients, thermodynamic observables, and screening quantities. In this sense, canonical forms of thermal integrands may become as central to future finite-temperature calculations as integration-by-parts identities have been over the past two decades.


\section*{Acknowledgements}

We acknowledge support from ANID under FONDECYT project No.~1231056 and Exploración Project No.~13250014.
All figures have been prepared with Axodraw \cite{Collins:2016aya}.


\end{document}